\def\kms{\ifmmode{\,\hbox{km}\,s^{-1}}\else {\rm\,km\,s$^{-1}$}\fi}

\def\msun{{\rm\,M_\odot}}
\def\lsun{{\rm\,L_\odot}}

\def\kmsm{{\rm\,km\,s^{-1}\,Mpc^{-1}}}

\def\hmpc{\ifmmode{h^{-1}\,\hbox{Mpc}}\else{$h^{-1}$\thinspace Mpc}\fi}

\def\eg{{\it e.g.}~}

\def\et{{\it et~al.}~}

\documentstyle[11pt,aaspp4]{article}
\tighten

\begin{document}

\title
{Galaxy Cluster Virial Masses and $\Omega$}

\author
{
R.~Carlberg\altaffilmark{1,2},
H.~K.~C.~Yee\altaffilmark{1,2},
E.~Ellingson\altaffilmark{1,3},
R.~Abraham\altaffilmark{1,4,5},
P.~Gravel\altaffilmark{1,2},
S.~Morris\altaffilmark{1,5},
\& C.~J.~Pritchet\altaffilmark{1,6}
}

\altaffiltext{1}{Visiting Astronomer, Canada--France--Hawaii Telescope,
	which is operated by the National Research Council of Canada,
	le Centre National de Recherche Scientifique, and the University of
	Hawaii.}
\altaffiltext{2}{Department of Astronomy, University of Toronto,
	Toronto ON, M5S~1A7 Canada}
\altaffiltext{3}{Center for Astrophysics \& Space Astronomy,
	University of Colorado, CO 80309, USA}
\altaffiltext{4}{Institute of Astronomy,
	Madingley Road, Cambridge CB3~OHA, UK}
\altaffiltext{5}{Dominion Astrophysical Observatory,
	Herzberg Institute of Astrophysics,
	National Research Council of Canada,
	5071 West Saanich Road,
	Victoria, BC, V8X~4M6, Canada}
\altaffiltext{6}{Department of Physics \& Astronomy,
	University of Victoria,
	Victoria, BC, V8W~3P6, Canada}


\begin{abstract}
The mean density of the universe is equal to the mass of a large
galaxy cluster divided by the equivalent co-moving volume in the field
from which that mass originated.  To re-examine the rich cluster
$\Omega$ value the CNOC Cluster Survey has observed 16 high X-ray
luminosity clusters in the redshift range 0.17 to 0.55, obtaining
approximately 2600 velocities in their fields.  The systemic redshift,
the RMS line-of-sight velocity dispersion, $\sigma_1$, and the mean
harmonic radius, $r_v$, are derived for each cluster using algorithms
which correct for interlopers in redshift space and measure the
angular extent of the sampling.  The virial mass, and its internal
error, are derived from these data. The cluster luminosity, corrected
to $z=0$, is estimated from the $r$ band luminosities of the cluster
galaxies.  Directly adding all the light to $M_r(0)=-18.5$, about
$0.2L_\ast$, and uniformly correcting for the light below the limit,
the average mass-to-light ratio of the clusters is
$283\pm27h\msun/\lsun$ and the average mass per galaxy is
$3.5\pm0.4\times10^{12}h^{-1}\msun$.  The clusters are consistent with
having a universal $M_v/L$ value (within the errors of about 20\%)
independent of their velocity dispersion, mean color of their
galaxies, blue galaxy content, redshift, or mean interior density.
Using field galaxies within the same data set, with the same
corrections, we find that the closure mass-to-light, $\rho_c/j$, is
$1160\pm130h\msun/\lsun$ and the closure mass per galaxy,
$\rho_c/\phi(>0.2L_\ast)$, is
$13.2\pm1.9\times10^{12}h^{-1}\msun$. Under the assumptions that the
galaxies are distributed like the mass and that the galaxy
luminosities and numbers are statistically conserved, which these data
indirectly support, $\Omega_0=0.20\pm0.04\pm0.09$ where the errors
are, respectively, the $1\sigma$ internal and an estimate of the
$1\sigma$ systematic error resulting from the luminosity
normalization.
\end{abstract}

\clearpage
\section{Introduction}

Clusters of galaxies are the largest collapsed objects in the universe
and play a particularly important role in the problem of estimating
the mean density of matter that participates in gravitational
clustering, $\rho_0$.  The standard procedure is as follows. The mass
of a cluster, within some radius, is measured with any of a number of
techniques, such as galaxy kinematics, X-ray profiles, or
gravitational lensing.  To relate cluster masses to $\rho_0=M/V$
requires an estimate of the co-moving volume, $V$, from which the
clusters collapsed, normally done via the luminosity of the cluster
galaxies, $L$, in ratio to the field luminosity density, $j$, so that
$V=L/j$. The value of $\Omega_0\equiv\rho_0/\rho_c$ can therefore be
rewritten as the cluster mass-to-light ratio, $M/L$, divided by the
closure mass-to-light ratio, $(M/L)_c\equiv\rho_c/j$.  The quantity
$\Omega_0$ is independent of $H_0$, and determines the future of the
expansion of the universe, in the absence of a cosmological constant
or a ``hot'' component of the mass field.  The classical cluster mass
estimator is the virial mass, $M_v$, a straightforward global
estimator suitable for relatively sparse data (but see \cite{bt81}).
This subject has a long history, with relatively stable results. That
is, the virial mass-to-light ratios of clusters, $M_v/L$, are
generally in the range $200-400h\msun/\lsun$ (\eg\
\cite{z33,smith,z37,schwarz,gunn78,rgh,david,neta95}) which, in ratio to
$(M/L)_c\simeq1500h$ (\cite{eep,loveday}) indicates
$\Omega_0\simeq0.2$.  The low value of $\Omega_0$ from cluster $M_v/L$
measurements is not generally accepted as a definitive measurement of
the field value of $\Omega_0$ because the random errors are generally
fairly large, and there are uncontrolled systematic errors in both the
mass and the light measurements which potentially allow the cluster
$M_v/L$ ratios to be consistent with $\Omega=1$.

There are two latent problems in the $M_v/L$ estimate of $\Omega$.
The reliability of the virial mass statistic as indicating the total
gravitational mass of a cluster is critically dependent on whether the
galaxy distribution traces the total mass distribution of the cluster.
The large color differences between cluster and field galaxies means
that their recent star formation histories are different, which may
simply be a recent decline associated with infall into the cluster,
or, it may indicate drastically different star formation efficiencies
at early times (\cite{dressler,dg,bo,cnoc_bo}) with the consequence
that the luminosity per unit mass could be quite different between
cluster and field. The possibility that the total mass distributions
of galaxy clusters are more extended than their constituent galaxies
has been recognized for many years (\cite{limber,wr,vbias}).  The
virial mass is a usefully accurate measurement of the mass within the
orbits of the galaxies, and completely independent of any anisotropy
of the velocity ellipsoid, but if the cluster light is more
concentrated than the cluster mass, the virial mass will be an
underestimate of the total mass.  This assumption of whether the mass
and galaxies are similarly distributed is straightforward to test
given sufficient data to derive radially resolved luminosity and
velocity dispersion profiles.

Dynamical mass measurements have the benefit that their strengths and
weaknesses are already relatively well understood and are easily
studied further with n-body simulations.  Although not of direct
interest here, dynamical mass estimation can be extended to well
beyond the virialized region, by measuring peculiar velocities in the
infall region, although then the results depend on the biasing of
galaxies with respect to the mass field.  The overall goal of this
project is to measure the total mass and luminosity contained within
the central virialized region of the cluster to establish a value of
$\Omega$, and measure any biases among cluster galaxies, cluster mass
and field galaxies.  Beyond the virialized region there is an
infalling mixture of galaxies, gas, and dark matter which is very
likely statistically identical to the field. Other than the few very
old central galaxies and a central cD the bulk of the cluster galaxy
population is likely to be composed of infallen field galaxies.
Because our sample of galaxies extends from the cluster core to the
distant field we can check whether galaxy population modifications
only occur in the virialized region of the cluster. Therefore we will
be able to empirically extend our measurement of $\Omega$ from the
virialized mass distribution to the $\Omega$ of the field. The scope
of this project lead to a broad collaboration under the title of
Canadian Network for Observational Cosmology (CNOC).

The twin complications for interpreting velocities are that cluster
galaxies and surrounding highly-correlated field galaxies are
intermingled in redshift space, and, individually clusters are quite
nonspherical, and internally clumpy, with the velocities at large
radii being influenced by surrounding large scale structure. Hence,
there are two main requirements for a successful measurement: accurate
control of the background and a sufficiently large sample to average
over the aspherical complications of single clusters.  A feasibility
study (\cite{cnoc1}) showed that approximately 1000 galaxy velocities
and a comparable number of field velocities distributed over 10 or so
cluster fields is sufficient to indicate whether the virial mass is
the total mass of the clusters. The built-in field sample allows a
study of the the relative galaxy populations of cluster and field, and
the closure mass-to-light and mass per galaxy.

There is no question that the cluster $M_v/L$ technique gives, in
principle, an accurate $\Omega$ value.  The key issues are to
constrain the technique's systematic errors and minimize its random
errors.  The main areas of concern are the accuracy of the virial
mass and galaxy population differences between the cluster and the
field.  Our data allow these problems to to addressed using our sample
alone, since it contains more than 1000 field galaxies, and offers
the considerable benefit of homogeneous selection as discussed in
Section 2.  In Section 3 the cluster RMS velocity dispersions and mean
harmonic radii are derived. These quantities are used to find virial
masses in Section 4. The luminosity and numbers of cluster galaxies
are used in Section 5 to give $M_v/L$ and $M_v/N_L$ ratios. The field
values of these quantities are derived in Section 6, and the resulting
$\Omega$ values in Section 7.  A companion paper (\cite{profiles})
will examine the relative light and mass profiles of the clusters.

\section{Sample and Observations}

Clusters at moderate redshifts, $z\simeq \onethird$, have a number of
advantages for mass estimation. They are sufficiently distant that
they have a significant redshift interval over which the density of
foreground and background galaxies are nearly uniformly sampled in
redshift. An accurate background estimate is crucial for measurements
in the outskirts of the cluster profile.  Clusters at higher
redshifts have increasing fractions of relatively blue
galaxies (\cite{bo}) which increases the similarity between
cluster galaxies and field galaxies (but they are by no means
identical), and, because the mean galaxy color varies, the effect of varying
galaxy types on the total luminosity can be measured.

There are a number of practical considerations which motivate our
choice of clusters.  At $z\simeq\onethird$, a comoving Abell diameter
of 3\hmpc\ spans an angle of about 13.3\arcmin, which is
sufficiently small that uniformity of photometry and sample selection
is relatively easily assured.  This size is also comparable to the
field size of the Canada-France-Hawaii Telescope (CFHT)
Multiple-Object-Spectrograph (MOS), approximately 10\arcmin\ square
(\cite{mos}).  These clusters are also ideal targets for other types
of mass measurement observations, such as gravitational lensing and
X-ray plasma profiles.

Our cluster sample was selected from the Einstein Medium Sensitivity
Survey Catalogue (\cite{emss1,emss2,gl}), which is a collection of
serendipitous objects discovered in X-ray images taken for other
purposes. For our survey we chose clusters with $z\ge 0.18$ (a
consequence of the poor blue response of the CCDs then
available) and in the
declination range $-15\le
\delta\le 65$, most suitable for CFHT. To select uniformly clusters that
are likely to be rich in galaxies, have a large velocity dispersion
(both of which make the measurement relatively easier) and are
guaranteed to have a substantial virialized component, we chose those
with $L_x\ge 4\times10^{44}$ erg~s$^{-1}$ and $f_x\ge
4\times10^{-13}$erg~cm$^{-2}$~s$^{-1}$. The A2390 cluster, a rich
(\cite{abellcat}), high X-ray luminosity (\cite{ulmer}) cluster was
added to fill an RA gap.

Observations were made at CFHT in 24 assigned nights in 1993 January,
June, October and 1994 January and March.  The observational
techniques and data reduction are described in \cite{yec} (hereafter
YEC), and the data will be described in a series of papers (\eg\
\cite{a2390_data}).  The fields (either EW or NS strips between
9\arcmin\ and 45\arcmin\ wide and about 8\arcmin\ high across the
cluster center) were imaged with the Multi-Object Spectrograph
(\cite{mos}) from which Gunn $g$ and $r$ magnitudes are derived, and
masks of spectrograph entrance slits were designed for the nonstellar
images.  The wavelength range of the spectra was shortened with band
limiting filters chosen to match the cluster's redshift, which
typically allowed about 100 targets to be observed on a single mask.
The limiting magnitude for spectroscopy for each cluster is set to
optimize the number of cluster (as opposed to field) redshifts
obtained. The resulting spectra are cross-correlated with a set of
templates to give velocities accurate to about 100 \kms\ in the rest
frame of the cluster.  The resulting ``pie diagrams'' for the entire
sample are shown in Figures~\ref{fig:raZ} and \ref{fig:dcZ}, where
redshift increases with distance from the vertex, and the angular
co-ordiante is the physical separation in the RA or Dec direction,
with the zero co-ordinate being at the location of the brightest
cluster galaxy (which is not necessarily at the center of the observed
field).

The fraction of galaxies for which we assign slits and obtain
redshifts decreases for fainter galaxies.  Such a selection function,
which decreases with magnitude, optimizes the rate of return of
cluster redshifts, but needs to be carefully measured to reconstruct
statistically the properties of the complete sample.  The algorithm
for deciding which galaxies will be assigned spectroscopic slits has
the highest priority at the expected $m_\ast$ of the cluster.  The
resulting $n(m)$ of the subsample with redshifts is nearly constant
with magnitude, whereas the total numbers rise with magnitude. The
redshift sample is corrected to the photometric catalogue with a
magnitude selection weight, $w_m(m_r)$. The limiting magnitude is
defined here as the $m_r$ where $w_m=5$. The MOS has approximately a
fixed number of slits per unit sky area (about 100 in 100 square
arcminutes), hence there is a geometric selection weight,
$w_g(m_r,x,y)$. For quantities that depend on surface area, these
weights need to be supplemented with the fraction of circle sampled by
galaxies at varying radii from the chosen centre. The defining
relations for these weights are described in YEC. All the results
given below incorporate these weights as is appropriate.

In this paper, all distance dependent quantities are calculated
assuming $H_0=100\kmsm$ and $q_0=0.1$, $\Lambda=0$. The choice of
$q_0$ is for approximate consistency with the most straightforward
interpretation of our results. The dependence of distance and
luminosity on $q_0$ largely cancels for the evaluation of $\Omega$
because it is defined relative to the field in the same redshift
range.  Our luminosities are corrected to a redshift zero for the sake
of approximate comparability with low redshift results.  These
corrected quantities should only be compared to others that our
adjusted to our photometric system. Hence, care should be exercised in
any direct comparison of our corrected $z\sim\onethird$ quantitites,
for instance mass-to-light ratios, to similar quantities at $z=0$.

\section{Cluster Dynamical Parameters}

The redshifts and positions of galaxies are used to define a
characteristic velocity and a characteristic length scale of the
cluster.  As a consequence of the virial theorem, the RMS velocity
dispersion has the important property that in a spherical system its
value is completely independent of any variation in the shape of the
velocity ellipsoid.  For a triaxial object the line-of-sight RMS velocity does
depend on viewing angle, which is precisely why we want to
average over a dozen or so clusters.
The line of sight velocity dispersion of a
cluster is defined as
\begin{equation}
\sigma_1^2 = \left({\sum_i w_i}\right)^{-1} \sum_i w_i (\Delta v_i)^2, \qquad
\label{eq:sig}
\end{equation}
where the $\Delta v_i=c(z_i-\overline{z})/(1+\overline{z})$ are the
peculiar velocities in the frame of the cluster and $\overline{z}$ is
the weighted mean redshift of the cluster. The weights used in
Eq.~\ref{eq:sig} are magnitude dependent geometric weights.
Unweighted velocities give similar results.

The virial mass estimator normally uses the projected mean harmonic
pointwise separation
\begin{equation}
R_H^{-1} = \left(\sum_i w_i\right)^{-2} \sum_{ij} {{w_iw_j}
	\over{\vert{\bf r}_i-{\bf r}_j}\vert},
\label{eq:rhc}
\end{equation}
where the $ij$ sum is over all pairs.  Being a pairwise quantity,
$R_H$ is sensitive to close pairs and is quite noisy
(\cite{bt81}). Furthermore there is no straightforward way to correct
$R_H$ for the approximately rectangular ``window'' which encloses our
cluster sample.  That is, simply weighting with the fraction of a
circular aperture that is enclosed within our rectangle leads to a
systematic underestimate of $R_H$.

Here we introduce an alternate estimate of $R_H$.  A pair of galaxies
at projected co-ordinates ${\bf r}_i$ and ${\bf r}_j$ are statistical
representatives of all galaxies at those projected radii.  Under the
assumption of axial symmetry the angle between these two vectors is a
uniform random variable. Hence, some immediate averaging is possible
by imagining that one (or the other) of the particles has its mass
distributed like a ring, with a radius equal to its radial location
with respect to some cluster center. More formally, the expectation
value for random angles of the pairwise potential $1/\vert{\bf
r_i}-{\bf r_j}\vert$ is the potential between a point and a ring,
which is (noting that the result is independent of the sign of the
cosine function for an integral over a circle)
\begin{eqnarray}
R_h^{-1}& =&\left(\sum_i w_i\right)^{-2} \sum_{ij}{w_iw_j{1\over{2\pi}}
\int_0^{2\pi} {d\theta\over\sqrt{r_i^2+r_j^2+2r_ir_j\cos{\theta}}}} \nonumber\\
&=&\left(\sum_i w_i\right)^{-2} \sum_{ij}{w_iw_j
	{2\over{\pi(r_i+r_j)}} K(k_{ij})},
\label{eq:rh}
\end{eqnarray}
where $k_{ij}^2=4r_ir_j/(r_i+r_j)^2$ and $K(k)$ is the complete
elliptic integral of the first kind in Legendre's notation
(\cite{nr}).  The quantity $R_h$ will be referred to as the ringwise
projected harmonic mean radius.  Unlike the original $R_H$ this
modified $R_h$ requires an explicit choice of the cluster center and
assumes that the cluster is symmetric about the center. Neither of
these pose any practical difficulties. The profile analysis will make
these assumptions, but it should be noted that for flattened clusters
the resulting $R_h$ will be a small overestimate if the field sampled
happens to lie along the major axis of the cluster.

There are two substantial benefits to be had from the definition of
$R_h$ in Equation~\ref{eq:rh}. For close pairs the divergence is
logarithmic instead of $1/r$, which makes $R_h$ less noisy than $R_H$.
Of immediate practical interest here is that the value of $R_h$ is
readily determined for data sets where a strip (sometimes of varying
width) across the center has been sampled. For instance, any of our
strips can be artificially narrowed to test how much $R_h$ varies. For
a factor of two reduction in the width of the A2390 data, $R_H$ declines
from 348 (already a substantial underestimate) to 267 arcseconds,
whereas $R_h$ drops from 533 to 517 arcseconds.

The three dimensional virial radius, $r_v$, is $r_v=\pi R_h/2$
(\cite{lm}).  With $\sigma_1$ and $r_v$ the virial mass of a cluster
is calculated as
\begin{equation}
M_v = {3\over G} \sigma_1^2 r_v.
\label{eq:virial}
\end{equation}

\subsection{Cluster Membership}

The complication in applying Equations~\ref{eq:sig} and \ref{eq:rh} to
redshift data is to decide which galaxies belong to the cluster and
which are likely to be interlopers in redshift space.  Because galaxy
groups and other clusters have such a high clustering probability near
a rich cluster this background is not expected to be smooth.  Our
magnitude distribution of galaxies with redshifts was designed to
maximize the abundance of cluster galaxies and minimize the
background. That is, the number of field galaxies per magnitude,
$n(m)$, rises steeply with magnitude, $n(m)\propto 0.4m$, whereas the
cluster luminosity function is much shallower, $n(m)\propto 0.25m$ or
less, below $m_\ast$, the characteristic brightness of a cluster
galaxy.  For each cluster we estimated $m_\ast$ and then set our
spectroscopic exposures so that we expected to obtain redshifts down
to the magnitude where cluster galaxies are starting to become less
numerous than field galaxies, which typically occurs 2.5 magnitudes
below $m_\ast$.  The higher priorities are assigned to galaxies with
brightnesses near $m_\ast$.  To meet the goals of this project it is
far better to have the same total number of cluster redshifts in many
clusters, rather than in a single one.  Besides being an efficient
procedure, this strategy also ensures that the cluster has the maximum
possible contrast in redshift space from the field, as shown in
Figures~\ref{fig:raZ} and \ref{fig:dcZ}. A sampling which is more
complete or deeper would mainly provide more field galaxies.  If more
velocities are desired it is far better to increase the sample of
clusters (to increase the statistical averaging over cluster to
cluster variations, such as substructure) rather than decreasing the
random error in a few clusters. However, fewer than 50 velocities in a
cluster often leads to large uncertainties because the velocity
structure of the cluster is not clear, in which case one tends to
overestimate the velocity dispersion.

The calculation of the RMS velocity dispersion of a cluster uses only
the redshifts with no positional information.  Defining the extent of
rich clusters in redshift space is a relatively difficult problem in
general (\eg\ \cite{bb,bird}) because cluster members moving at a
large velocity with respect to the center of the cluster are
impossible to distinguish from field galaxies.  Furthermore, many
statistical tests to assess the membership status of galaxies in the
tails of the velocity distribution are based on the assumption that
the underlying distribution is a unimodal Gaussian, which is generally
not true for such a kinematically complex system as an accreting
cluster and the surrounding large scale structure. We have reduced the
problem through the magnitude distribution of our cluster sample,
which provides data that leaves clusters as cleanly defined as
possible in redshift space.

The procedure for defining the redshift range of the cluster is a
manually iterated procedure, roughly equivalent to measuring the width
of an emission line in a spectrum. First, an initial estimate of the
outer limits of the redshift range of the cluster is made, using
Figures~\ref{fig:raZ}, \ref{fig:dcZ}, and \ref{fig:zb}. The redshift
limits adopted are shown as dotted lines in Figure~\ref{fig:zb}. The
initial estimate of the redshift range is used to calculate a trial
velocity dispersion, $\sigma_1$.  A second algorithm is used to assess
the validity of this velocity dispersion in the presence of
interlopers in the outer parts of the redshift space of the cluster.
The velocity dispersion validation algorithm works as follows.  The
weighted (as in Eq.~\ref{eq:sig}) velocities within $15\sigma_1$ (or
the edge of the band-limiting filter if that is smaller) of the
cluster center are put into bins $0.1\sigma_1$ wide (within reason the
bin width has no effect).  The mean density of the background in
velocity space is calculated from the mean bin density $>5\sigma_1$
away from the cluster.  This background is then subtracted from the
data within $3\sigma_1$ of the cluster center and the velocity
dispersion $\sigma_1^\prime$ is calculated, along with a Bootstrap
estimate of the error in $\sigma_1^\prime$. If $\sigma_1^\prime$ and
$\sigma_1$ are within one standard deviation, $\sigma_1$, is accepted
as the cluster's velocity dispersion; if they are not equal, the
redshift limits are further adjusted.

The adopted ranges in redshift, as defined by the above procedure, and
radius, as defined by the rectangular region which bounds the data,
are given in Table~\ref{tab:lim}. Column 2 gives the weighted mean
redshift, columns 3 and 4 gives the dimensions of the bounding box in
RA and Dec as converted to physical lengths, $D_A(z)\Delta\theta$, at
the redshift of the cluster.  Columns 5 and 6 give the lower and upper
redshift boundaries.  In Table~\ref{tab:sig} the second column gives
the ratio of the ``unlimited'' $\sigma_1^\prime$ to the value
accepted, and the third column gives the 1 standard deviation
confidence interval for the $\sigma_1^\prime/\sigma_1$ ratio.  Each
cluster's velocity dispersion is calculated using
Equation~\ref{eq:sig} with the weighted data in the cluster's redshift
range given in Table~\ref{tab:lim}. The results are given in
Table~\ref{tab:rsd}. Column 2 of Table~\ref{tab:rsd} is the redshift,
column 3 is the number of galaxies above the magnitude limit in the
cluster RA, Dec and redshift range. Column 6 is $\sigma_1$. The sky
positions of the cluster galaxies for the inner 1000 arcsec of the
sample are shown in Figure~\ref{fig:cL}.

The obvious redshift limits were acceptable in about half of the
clusters.  In the other half the limits were adjusted until a
statistically acceptable result was found, or until no stable value
emerged.  The velocity dispersion checking algorithm indicates
(Table~\ref{tab:sig}) that the velocity dispersion of MS0906+11 is
significantly too large.  No consistent solution could be found for
this cluster, which appears to be an indistinct binary in redshift
space.  All other clusters have velocity dispersions that are well
within two standard deviations of the input values.  Based on this
test, we will accept the results of Table~\ref{tab:rsd} as our
standard values. One could apply the slight corrections of
Table~\ref{tab:sig}, but that is simply changing the values within
their standard error. The cluster MS1358+62 appears to be an unequal
binary in declination (Fig.~\ref{fig:dcZ}), although its velocity
dispersion appears to be acceptable.

The modified mean harmonic radii, $R_h$, are calculated for each
cluster using the entire angular extent of the available data in the
redshift range of the cluster as fixed for the $\sigma_1$ calculation.
The mean interior overdensity inside $r_v$ is an indicator of the
radial range of the cluster that has been observed.  The angular
extent of the clusters is defined by the angular limits on the sky of
our sample.

It is expected (\cite{gg}) and borne out by numerical simulation (\eg\
\cite{cer}) that, on the average,
substantially all the virialized cluster mass is contained inside the
radius where $\overline{\rho}>200\rho_c$.  Outside that radius the
velocities very rapidly become dominated by infall.  Therefore the
ratio $\overline{\rho}(r_v)/\rho_c$ is used as a test of the
completeness of the radial extent of the sampling of the cluster.
This ratio is calculated as the mean density inside $r_v$ to the
critical density at the observed redshift by dividing the virial mass
(Eq.~\ref{eq:virial}) by the volume inside $r_v$,
\begin{eqnarray}
{\overline{\rho}(r_v)\over{\rho_c(z)}} & = & {1\over{\rho_c(z)}}
	        {{3 M_v}\over{4\pi r_v^3}} \nonumber \\
&= &{6\over{(1+z)^2(1+\Omega_0 z)}}
        {{\sigma_1^2}\over{H_0^2 r_v^2}}.
\label{eq:rho}
\end{eqnarray}
The critical density at redshift $z$ is
$\rho_c(z)=\rho_c(0)(1+z)^2(1+\Omega_0 z)$, which for $\Omega_0=1$, has
the $(1+z)^3$ dependence of the mean density.  For $\Omega_0<1$
Eq.~\ref{eq:rho} means that the density ratio at $z$ is somewhat
larger than for $\Omega_0=1$, with the difference being 25\% at
$z=\onethird$ for our adopted $2q_0=\Omega_0=0.2$.

The {\em total} virialized mass of the cluster should be
accurately estimated if the mean density inside $r_v$ is $200\rho_c$
or less.  Table~\ref{tab:rsd} includes 6 clusters for which this
criterion is met.  For reference, the Coma cluster has $R_h\simeq
1.78$\hmpc\ (1.48$^\circ$, \cite{schwarz}) and, with $\sigma_1$ of 1040
\kms\ (\cite{peebles_coma}) has $\overline{\rho}(r_v)= 77\rho_c$.

\subsection{Error Analysis}

The errors in $\sigma_1$, $R_h$, $M_v$, and $M_v/L$ are assessed using the
Jacknife technique. This is one of the simplest resampling techniques,
wherein partial standard deviations, $\delta_i$, are calculated by
taking the difference between the $f(x_1,\ldots,x_n)$, where $f$ is
any statistical quantity calculated from the entire data set, and the
same quantity calculated dropping one element of the data set,
$\delta_i =
f(x_1,\ldots,x_n)-f(x_1,\ldots,x_{i-1},x_{i+1},\ldots,x_n)$.  The
estimate of the variance is $\sqrt{n/(n-1)\sum_i \delta_i^2}$
(\cite{efron,et}).  For a Gaussian distributed sample the Jacknife
variance converges to the normal value.  A closely related technique
is the Bootstrap, in which randomly drawing {\em with replacement}
from the original data set creates many new data sets of the same
size as the original which are then analyzed in precisely the same
manner as the original to give a distribution of results from which
confidence intervals are calculated. A significant difference is that the
Bootstrap data sets have repeated values, which for pairwise
statistics gives singularities. Also, the Bootstrap can be
computationally quite expensive, since tests indicate (\cite{et}) that
300 or more resamplings are required to give solid results.  The
errors given in Table~\ref{tab:rsd} (in the columns labeled
$\epsilon_r$ and $\epsilon_\sigma$) are approximately in accord with a
$\sqrt{N}$ expectation, but have significant differences in
detail. The reliability of these errors is extremely important for one
of the major results of this paper. We believe the pragmatic nature of
the Jacknife error estimate is well suited to the task.

\subsection{Sample Commentary}

A cluster-by-cluster comparison with the results of other workers is
done in the data papers for each cluster.  In a number of cases our
velocity dispersions are notably lower than those previously found.
For instance, we find for Abell 2390 that $\sigma_1=1100\pm63$\kms\
whereas a previous study in the central region found
$2112^{+274}_{-197}$\kms (\cite{leborgne}).  In general the
differences from previous work can be attributed to three factors.
First, the precision of our velocities, $\simeq$100 \kms, reduces
catastrophic velocity errors and helps identify significant
substructures.  Second, our data cover a large radial range in the
cluster, which makes them less subject to local substructure, and more
representative of the RMS value.  Third, the bluer galaxies, which
often contain measurable emission lines, statistically are found to
have a higher velocity dispersion than the redder absorption line
galaxies, an effect which is particularly prominent near the projected
center of the cluster (\cite{a2390,profiles}). Since redshifts are
easily measured from emission lines this can lead to an upward bias in
the velocity dispersion, but which is demonstrated to be quite small
in our data (\cite{yec}).  These effects can be deadly in the presence
of substructure, for which there is evidence in many of our clusters.
In general the clusters are quite regular, possibly connected to their
high X-ray luminosity, and usually does not upset any of the results
below. The exceptions are identified and removed, as discussed below.

The quantity $\overline{\rho}(r_v)$ in Table~\ref{tab:rsd} varies from
a low of 84$\rho_c$ in MS1621+26 to 1135$\rho_c$ in MS1455+22,
discounting MS0906+11.  It is clear that $\overline{\rho}(r_v)$ is
high for those clusters for which only a central field was observed.
In particular, we expect on the basis of Coma data and n-body
simulations, that it is necessary (but not sufficient) that
$\overline{\rho}(r_v)$ should be less than $200\rho_c$ if the formal
$M_v$ (Eq.~\ref{eq:virial}) is interpreted as the total virialized
mass.  If $\overline{\rho}(r_v)>200\rho_c$ then it indicates that
$r_v$ is an underestimate of the true value for the entire virialized
system, which is a straightforward consequence of the limited angular
sampling of some of our clusters.  This undersampling is an important
issue, but is not a significant problem for the major goals of this paper
because some quantities, such as $M_v/L$, are insensitive to the outer
limits of the sampling, and it is possible to define a characteristic
radius at a fixed mean interior density, which is insensitive, within
reasonable limits, to these sampling variations.

\section{Virial Mass Analysis}

The quantities $\sigma_1$ and $r_v$ are given in Table~\ref{tab:rsd}.
The resulting $M_v$ (Eq.~\ref{eq:virial}) are given in
Table~\ref{tab:ml}, where column 2 repeats the $\overline{\rho}(r_v)$
column of Table~\ref{tab:rsd}, column 3 is $M_v$ in units of solar
masses, column 4 is the total observed luminosity in units of solar
luminosities (see Section 5), columns 5 and 6 are the canonical $M_v/L$
ratio and its standard error from the Jacknife technique.  It must be
borne in mind that the masses given are defined by the radial extent
of the sample, and must be referred to the mean density given in
column 2.  That is, if the cluster extends beyond the size of our
observed field, our mass is an underestimate of the total
mass. Nevertheless, it remains of considerable interest to compute an
$M_v/L$ for the part of the cluster we observe. Furthermore it is
relatively straightforward to bring all the masses to a common ground.

The virial mass, as calculated from a data set which misses the outer
parts of the cluster, will still accurately estimate the mass contained
within the orbits of the galaxies observed, with a weak dependence on
the radial extent of the sampling as we show here.  The virial theorem
is a derived by multiplying the Jeans equation of stellar
``hydrostatic'' equilibrium (a vector equation) with co-ordinate
vectors and integrating over the volume of the system
(\eg\cite{bt_gd}). The outer limit of the volume integral is usually
taken at infinity, but, when evaluated for a finite subset of the
entire volume, gives an additional term to be evaluated at the outer
surface.  The scalar virial theorem for a spherically symmetric system
bounded at radius $r_b$ becomes
\begin{equation}
\int_0^{r_b} (\sigma_r^2+2\sigma_\theta^2)\rho\, dV
+ \int_0^{r_b} {{-GM(r)}\over{r}}\rho\, dV = 4\pi\rho\sigma_r^2 r_b^3.
\label{eq:svir}
\end{equation}
The right hand side of Eq.~\ref{eq:svir} is sometimes called the $3PV$
surface term. The effect of ignoring the surface term for a
gravitating system is to overestimate the mass of the system, since
the surface pressure reduces the amount of mass to keep the system in
equilibrium.

A singular isothermal sphere gives an estimate of the largest
reasonable error if we use Eq.~\ref{eq:virial} on a partially sampled
cluster.  The isothermal sphere provides an upper limit, since
equilibrium stellar systems normally have a falling velocity
dispersion with radius, so the isothermal sphere has an unusually
large surface term.  The density profile of a singular, isotropic,
isothermal sphere of velocity dispersion $\sigma=\sigma_r$ is
$\rho=(2\pi G)^{-1}\sigma^2 r^{-2}$.  Equation~\ref{eq:svir} then
becomes
\begin{equation}
3\sigma^2M_b  -{{GM_b^2}\over r_b} = \sigma^2 M_b,
\label{eq:vir_surf}
\end{equation}
where $M_b$ is the mass inside the boundary.  If the surface term is
neglected, then the derived virial mass will be an overestimate of the
mass contained inside $r_b$, $3G^{-1}\sigma^2 r_b$, instead of the
correct $2G^{-1}\sigma^2 r_b$.  If the density profile declines more
rapidly than $r^{-2}$, as is indeed the case for the clusters studied
here, then the error introduced through the neglect of the surface
term is reduced. The isothermal sphere has the pleasing property that
the relative size of the surface term is completely independent of
shape or size of the surface.  We conclude that if only an inner part
of the cluster profile is sampled, then the calculation of the virial
mass will be at most a 50\% overestimate.

Under the assumption that clusters are approximately $\rho(r)\propto
r^{-2}$ (or any other mass profile of interest), the masses can then
be cautiously extrapolated to other radii or mean interior
densities. The results of extrapolating the masses to a constant mean
interior density of $200\rho_c$ using
$M_{200}=M_v\sqrt{\overline{\rho}(r_v)/(200\rho_c)}$ are given in the
second last column of Table~\ref{tab:ml}.

In ``raw'' $M_v$ the most massive cluster in the sample is A2390
(discounting MS0906+11), with $M_v=1.7\times10^{15}h^{-1}\msun$, which makes
it comparable to Coma's $2.1\times 10^{15}h^{-1}\msun$. Coma and A2390
have similar mean interior overdensities at $r_v$, 77 and 110,
respectively, putting the mass comparison on nearly the same density
basis.  The masses of the other clusters range a factor of about 2.5 bigger
and smaller than $M_{200}\sim 5\times10^{14}\msun$. This rather narrow
range in mass is a consequence of our X-ray luminosity and flux
selection.  It is perhaps not too surprising that one of the most
massive clusters, MS0016+16, is also the highest redshift one in our
sample.

\section{Cluster Galaxy Luminosities and Numbers}

There are two practical approaches to relate empirically the cluster
contents to a co-moving volume in the field. A third approach, a total
cluster baryon inventory (\cite{wnef}), cannot at present directly
relate the cluster to the field since most field baryons are in some
unobserved form and one must turn to a Big Bang Nucleosynthesis
argument (\cite{walker}) to obtain a field density of the baryons.
However, the large reservoir of non-stellar baryons in clusters raises
concern about the assumption that the fractional conversion of gas
into stars has been the same in clusters and the field. Our galaxy
sample extends from cluster to the far field, allowing a number of
tests for variation. The two ``binary'' clusters, MS0906+11 and
MS1358+62 are excluded from all averages calculated below.

The total luminosity of a cluster estimates the co-moving volume from
which it collapsed, $V=L/j$. This approach assumes that the total
luminosity is conserved, or that any systematic change in galaxy
luminosities during the gravitational assembly of the cluster can be
measured and corrected. However the subject of galaxy evolution within
clusters remains quite controversial, at the very least because the
observational phenomena are not yet well defined.

An alternate approach to measuring the equivalent field co-moving
volume of the cluster is to use the numbers of cluster galaxies above
some absolute magnitude in ratio to the field,
$V=\rho_c/\phi(>L)$. Each of these methods has different strengths and
weaknesses, but they are substantially {\em independent} since cluster
total luminosities are dominated by the brightest galaxies but the
total numbers are dominated by the low luminosity galaxies.  One
possible evolutionary effect is the merging of the stellar components
of two galaxies which makes no difference to the total luminosity if
there is no accompanying episode of star formation. Merging does, of
course, decrease the numbers of galaxies.  It is well known that
cluster galaxies are on the average significantly redder than field
galaxies (Figure~\ref{fig:light}), which is taken to indicate that
star formation has decreased in cluster galaxies and hence they are
less luminous per unit stellar mass than in the field.  Fading will
decrease the numbers of cluster galaxies above some fixed luminosity
limit relative to the field, although, if the limit is well below
$M_\ast$, then the decreased numbers will be a small fraction of the
change in luminosity. For instance, if all cluster galaxies down to
$M_r(0)=-18.5$ were included in the total count, but there had been a
one magnitude fading with respect to the field, then galaxies down to
$-17.5$ should have been included in the total count. The expected
increase in numbers (for a Schechter function with $M_\ast=-20.3$ and
$\alpha=1.25$) would be only 17\%, as compared to a 150\% correction
required to the luminosity.  Cluster galaxies are not simply post-star
formation field galaxies with diminished luminosities, since clusters
contain a population of very luminous, but very red, Elliptical and cD
galaxies in their centers. These may have been created through
mergers, or they could have tapped the large baryon reservoir present
in the cluster to create more stars. If galaxy luminosities are either
increased or decreased, the $L$ of a cluster responds in proportion,
but the numbers of cluster galaxies above some magnitude change far
less. The evolution of cluster galaxies relative to field galaxies,
especially at these redshifts, is sufficiently ill defined that we do
not, at this stage, want to make a highly uncertain differential
correction.  However, we will use mass-to-light and mass-per-galaxy to
obtain two substantially independent estimates of $\Omega$, which in
turn will be used to address the notoriously difficult problem of
constraining the size of systematic errors. That is, if the two
estimators give discrepant values beyond the errors, then we would
have detected a systematic effect (whose source would then be
identified and corrected, if possible). We cannot empirically detect
systematic errors smaller than our random errors which then becomes
our systematic error estimate.

The observed luminosity of the cluster galaxies, summed to
$M_r(0)=-18.5$, corrected as described below, is reported in
Table~\ref{tab:ml}. The luminosity is measured in the Gunn $r$ band,
using $M_r(\odot)=4.83$. We also note that $r$ band selection has the
useful benefit that it is not unduly sensitive to star formation
increases, although these are readily detected in the $g-r$
colors. The measured $m_r$ are K corrected using the following
method. Model K-corrections in $r$ and $g-r$ colors as a function of
redshift for non-evolving galaxies of 4 spectral types (E+S0, Sbc,
Scd, and Im) are derived by convolving filter response functions with
spectral energy distributions from \cite{cww}. These values are then
corrected from the AB system to the standard Gunn system (\cite{tg}).
For each galaxy with redshift, a spectral classification is estimated
by comparing the observed $g-r$ color with the model colors at the
same redshift.  The spectral classification, obtained via
intepolation, is treated as a continuous variable within the 4
spectral types.  From the spectral classification, the appropriate
K-correction to the $r$ mangitude is then derived using the models.
Our data strongly indicates that the K corrected cluster luminosity
function is brightening with redshift (\cite{yec2}), hence our
luminosities are given a mild evolutionary correction, $E(z)= -z$
(\cite{yg}), so that they are appropriately faded to the current
epoch.

The corrected magnitudes and colors are
shown in Figure~\ref{fig:light}.  An entirely empirical redshift
normalization of the colors, $(g-r)_z = (g-r)/(1.2+2.33*(z-0.3))$, is
shown.  This relation is based on a linear fit to the redshift
dependence of the red sequence of the bright cluster galaxies.
Because the median color of these clusters is always quite red, there
is little range in the quantity $(g-r)_z$.  The fraction of the
galaxies that are blue can be estimated with a parameter similar, but
quite different in detail, to the Butcher-Oemler blue fraction
(\cite{bo}).  The quantity $F_b$ here is the fractional weight of all
galaxies bluer than $(g-r)_z=0.7$. Note that this includes all
galaxies used in the total luminosity and does not restrict the
measurement of the blue fraction to some fiducial region. A more
detailed analysis of the Butcher-Oemler effect will be presented
elsewhere (a preliminary account is in \cite{cnoc_bo}).

It is conventional to quote $M_v/L$ values which are based on the
``total light'', that is, a correction is made for the part of the
luminosity function below the limiting magnitude.  This is not
strictly necessary here, since the field sample is treated in
precisely the same manner, but for the sake of comparability we do
this correction, for both the cluster and field luminosities.  The
``unobserved light'' beyond the limiting magnitude our sample is
estimated assuming $M_\ast=-20.3+5\log{h}$ and $\alpha=-1.25$ in a
Schechter luminosity function. A useful limiting magnitude is about
$M_r= -18.5$, which is about $0.2L_\ast$. The figures and tables of
this paper use this limiting luminosity, unless otherwise stated,
which helps put all the $M_v/L$ values on the same basis.  There is
very little difference in the estimated $L$'s if we include all the
galaxies to the sampling limit.  The $\Omega$ estimate from the
$M_v/L$ has no dependence on the correction to the total luminosity,
since the field galaxies will have precisely the same multiplicative
correction included.

The luminosities, like the masses, are the total luminosities for the
sample, and must be referred to its mean interior density. The
luminosities can be extrapolated to a common mean density, $L_{200}$,
in the same manner as the masses.  That is,
$L_{200}=L\sqrt{\overline{\rho}(r_v)/(200\rho_c)}$, which builds in a
correlation between $L_{200}$ and $\sigma_1^{1/3}$, but not with the
dynamical range observed here.  The cluster quantities are
defined on the basis of a constant ratio with respect to
the critical density at the redshift of the cluster, $\rho_c(z)$,
which increases with redshift. Singe $\sigma^2\propto \rho^{1/3} M^{2/3}$,
with the redshift dependence of $\rho$ included this becomes
$\sigma_1\propto(1+z)^{1/3}(1+\Omega_0z)^{1/6}M^{1/3}$,
Figure~\ref{fig:ls} plots the redshift adjusted luminosity against
the velocity dispersion.  This plot excludes MS0906+11 and
MS1358+62.

\subsection{Cluster Mass-to-Light Ratios}

It is immediately apparent from Table~\ref{tab:ml} that the $M_v/L$ do
not show much variation outside their errors, in spite of the
variations in the sampling radii.  The mean $M_v/L$ for the sample
(excluding MS0906+11 and MS1358+62) is $283\pm27h\msun/\lsun$ (in Gunn
$r$, corrected to $z=0$ as described above), and $245\pm32h\msun/\lsun$
for the six clusters that have a mean interior density less than
$200\rho_c$, excluding the asymmetric MS1358+62 cluster.  The $M_v/L$
values as a function of redshift are shown in Figure~\ref{fig:ml}.  It
should be noted that luminosity evolution has been included, at the
rate of $\Delta M=-z$. In the absence of this correction the
luminosities at $z=0.55$ would be 1.66 times higher, and would lead to
a noticeable gradient in the diagram. The corrections to the
luminosities are all multiplicative factors that are applied both to
the field and the cluster data, so they have no effect on the $\Omega$
value derived.

The distribution of $M_v/L$ differences from the 14 cluster sample
mean, normalized to their measurement errors, is also displayed in
Figure~\ref{fig:ml}.  The cluster at $-3.6$ standard deviations,
MS1621+26, is one of the clusters with the best data and is also one
of the bluest clusters, possibly indicates that its luminosity is
higher than it would be if the galaxies were allowed to age enough
that the colors became comparable to other clusters in this redshift
range.  The redshift-$M_v/L$ plot of Figure~\ref{fig:ml} also gives
the impression that all the clusters have the same $M_v/L$ (removing
MS0906+11 and MS1358+62) within the errors.  The $\chi^2$ per
degree-of-freedom is 1.8, but removing MS1621+26 as well reduces it to
1.03 which is consistent with no variation beyond the errors.  That
is, the $M_v/L$ of Table~\ref{tab:ml} are consistent with a universal
underlying $M_v/L$, after a 3 standard deviation clipping is applied.
This means that the measured cluster light is a very good indicator of
the mass contained within the orbits of the galaxies. It is intriguing
to note that this appears to be true over a substantial range of mean
interior densities which characterize the cluster samples. A similar
result has been found for X-ray measurement of cluster masses
(\cite{david}), although their $M/L$ profiles values decline with
increasing overdensity, whereas our integrated values are nearly
constant, or rise slightly, with mean interior overdensity, within the
limits expected from the virial surface term of Eq.~\ref{eq:vir_surf}.

The percentage errors in the $M_v/L$ values are in the 10-30\% range,
depending on the number of cluster galaxies. With more data it seems
quite likely that it will become evident that there are intrinsic
variations in cluster $M_v/L$ values. In particular the cluster light
could have a ``second order'' luminosity correction which would depend
on the precise mix of galaxies in the cluster.  This is already the
case with MS1621+26 which has 98 members with redshifts and is
somewhat ``blue'' compared to the trend of colors with redshift.  Of
course any true $M_v/L$ variations cannot be bigger than the 20-30\%
error bounds found here.

\subsection{Cluster Mass-to-Number Ratios}

The mass per galaxy for all galaxies brighter than some chosen
absolute magnitude is less sensitive to differential fading (or
possible brightening) between the field and cluster than the mass per
unit luminosity. Merging could alter the galaxy numbers, but, in as
much as mergers lead to ellipticals (the E+A galaxies appear to be
mostly disk systems, \cite{dobg}), it does not drastically alter the
numbers of galaxies.  Both fading and merging will cause the $M_v/N_L$
ratio to overestimate the value of $\Omega$.  The individual
mass-to-number ratios to a limiting magnitude of $M_r(0)=-18.5$ are
shown in Figure~\ref{fig:mn} and given to $-18.5$ and $-19.5$ in
Table~\ref{tab:mn}.  The $\chi^2$ for a limit of $-18.5$ is very large,
about 3 per degree of freedom.  At a limit of $-19.5$ the $\chi^2$ per
degree of freedom drops to 0.6 if MS1621+ 26 is dropped.

Like $M_v/L$, $M_v/N$ shows no significant dependence on other cluster
parameters.  Figure~\ref{fig:mn} weakly suggests an increase in
scatter as a function of the mean interior overdensity, although this
is likely to be primarily a statistical effect, since the highest
density clusters are those that are least well sampled.  There is a
tendency for the high velocity dispersion clusters to have the largest
$M/N_L$ values. Although this trend is not very significant, it is
opposite to a standard biasing scheme, ``peak-background'', in which
galaxy formation is initiated earlier and is more advanced in clusters
of increasing mass.

The average values of $M/N_L$ for all the clusters are given for a
range of limiting magnitudes in Table~\ref{tab:mlnav}.  The table
demonstrates that the $M/N_L$ is not very sensitive to the limiting
absolute magnitude if it is at least 2 magnitudes less than $M_\ast$,
but that the sensitivity rises quickly (exponentially is expected for
a Schechter luminosity function) near $M_\ast$.  The reduced $\chi^2$
of the deviations from the mean are very large for low luminosity
limits. In the case of a limit of $M_r(0)=-19.5$ the reduced $\chi^2$
drops to 0.8.  This is partially a consequence of the smaller number
of cluster galaxies, typically 60\% of the full sample, leading to an
increase in the random errors. None the less, the reduction in
$\chi^2$ is so striking that it suggests that the most luminous
cluster galaxies have far less cluster-to-cluster variation than the
lower luminosity ones, which is a comment that has been made in other
contexts.

\section{Light and Number Densities of Field Galaxies}

The closure mass-to-light ratio is defined as $(M/L)_c\equiv\rho_c/j$,
where $\rho_c=3H_0^2/(8\pi G)$, and $j=\int_0^\infty L\phi(L) \,dL$,
where $\phi(L)$ is the field luminosity function. The integration is
carried to low luminosities by assuming a Schechter form for the
luminosity function for galaxies below the absolute magnitude cutoff.
Similarly the closure $(M/N_L)_c\equiv\rho_c/\phi(>L)$, where $N_L$ is
the number of galaxies more luminous than $L$.  At low redshift
$(M/L)_c$ is estimated to be about $1500^{+700}_{-400}h\msun/\lsun$
(\cite{eep,loveday}).  This value is based on a blue selected sample,
and integrates all the light in the luminosity function to very faint
magnitudes. Our selection is based on Gunn $r$, and we do not sample
galaxies much fainter than $M_r\simeq -18.5$.  These differences would
lead to substantial systematic errors in $\Omega$ if we tried to ratio
moderate redshift cluster light to a low redshift field estimate of
the closure density.

Our survey includes a built-in field sample which can be used to
estimate accurately $(M/L)_c$ and $(M/N)_c$, on the same manner as our
cluster values. Each field galaxy is K corrected and evolution
corrected, and within a redshift range we correct the total luminosity
for light below our magnitude limit in precisely the same as the
cluster galaxies. Because all of these factors are multiplicative,
they cancel in the eventual ratio of interest for $\Omega$.  The
corrections are done for the sake of providing numerical values having
some degree of comparability to true low redshift quantities. The
luminosity density of the field can be nearly trivially calculated
from our field sample, and offers the benefits that it is selected in
precisely the same way, in the same filter band, in the same redshift
range, and to the same limiting magnitude as the cluster data.  This
approach, using a self-consistent luminosity system for the cluster
and field, greatly reduces concerns about systematic differences
between the low redshift estimate and our medium redshift data.

Table~\ref{tab:close} gives the results of the $(M/L)_c$ and
$(M/N_L)_c$ calculation to the indicated limiting magnitudes, averaged
over the redshift range $0.2\le z \le 0.6$. The cluster redshift
ranges of Table~\ref{tab:lim} are widened by $\Delta z=0.01$ at both
the upper and lower redshift to eliminate any concerns about an
overestimate of $j$ from a higher density of galaxies in the vicinity
of these very rich clusters. In fact there is no significant change in
the field luminosity density without this extra cut around the
clusters. The weighted luminosities of the field galaxies, corrected
to $z=0$ precisely as the cluster data, are simply summed in each
redshift bin, cutting the sum off at various absolute magnitudes.  The
volumes are in {\em co-moving} $\hmpc^3$ integrated over the
accessible redshift range of the bin for our chosen $q_0$, times the
total solid angle of our survey, 2367 square arcminutes.  There is no
statistically significant redshift trend in either the luminosity
or number density in our evolution corrected magnitude system. If
evolution correction were not included the values would be about 35\%
smaller, and an evolutionary trend would be marginally significant in
the luminosity density.

The random errors of our closure $(M/L)_c$ (column 4 of
Table~\ref{tab:close}) are calculated using the Jacknife technique
assuming that each cluster field should be treated as a data
element. This might be a conservative estimate of the errors, but fully
accounts for large scale structure variations.

The colors of the field galaxies are shown in Figure~\ref{fig:light}.
The field galaxies are always substantially bluer than the cluster
galaxies: the average $(g-r)_z$ is 0.55, with no significant change
with redshift over the entire range of $0.1\le z \le 0.6$. The
cluster galaxies have average $(g-r)_z$ values ranging from 0.72 to
0.94. Although there is some overlap of the populations, there is no
overlap of the averages.  Hence it is not possible to empirically
estimate from this sample the luminosities of the cluster galaxies if
they had the same colors as the field galaxies.

\section{$\Omega$}

The value of $\Omega$ is estimated under the assumption that the
$M_v/L$ of the clusters is the same as $M/L$ in the field.  Our best
estimate, in the sense of smallest random errors is the $M_v/L$
estimate to a limiting magnitude of $-18.5$ for which we find that
$\Omega=0.25\pm0.05$ at a mean redshift of 0.31. Brighter absolute
magnitude limits give consistent values of $\Omega$, but with
increased errors.

The value of $\Omega$ from the $M_v/N_L$ calculation is $0.28\pm0.08$,
for absolute magnitude limits of both $M_r(0)=-18.5$ and $-19.5$.
This value is consistent with the one found from the $M_v/L$ argument.
We will prefer the $M_v/L$ value of $\Omega$ mainly because of its
smaller errors, but also as the conventional approach.  The fact that
the $M_v/L$ and the $M_v/N_L$ routes to calculating $\Omega$ gives
similar results is considerable grounds for confidence in the
result. The two $\Omega$ values have substantially different
dependencies on the evolution of cluster galaxies, and are weighted to
different parts of the luminosity function.  Certainly we would not be
able to detect any systematic errors in cluster galaxy evolution which
is smaller than the quadrature summed random errors of the two
$\Omega$ results.  We will use this error sum as our estimate of the
likely size of any residual systematic errors in our cluster $\Omega$
values, due to galaxy evolution alone. It does not measure any
systematic dynamical errors in the masses.

Tables~\ref{tab:mlnav} and \ref{tab:close} both have a a common trend
$M/L$ values where fainter imposed cutoffs, which use more of the real
data, find less light than the adopted luminosity function does,
although the trend is not very significant.

Since $\Omega_0=\Omega_z/(1+z(1-\Omega_z))$, for our average redshift
of 0.31 this is $\Omega_0=0.20\pm0.04$, where the formal 1 standard
deviation random error is given.  The strength of this value is that
is entirely from our survey alone, with its well understood sampling
statistics, requires no low redshift comparisons, and is done entirely
within one photometric system which requires only small redshift
corrections.  The main issue which is addressed in the companion paper
(\cite{profiles}) is whether the virial mass correctly estimates the
total mass of clusters, with further examination of the differences
between cluster and field light per unit mass.

\section{Conclusions}

The CNOC cluster sample is a uniform, X-ray selected, sample of
clusters having $0.17<z<0.55$. We derive velocity dispersions and a
characteristic radius for each of these clusters, and calculate their
virial mass, based upon objective tests to estimate the radial and
redshift extent of the cluster members.  The Gunn $r$ band luminosity
of the clusters is corrected to z=0, and the ratio $M_v/L$ to a
limiting magnitude of $-18.5$ (corrected to a total luminosity with a
Schechter luminosity function), is found to have a mean value of
$283\pm27h\msun/\lsun$, comparable to low redshift clusters.  The
clusters are consistent with having identical $M_v/L$ values (for the
corrected luminosities) within the sample variance of
$\pm100\msun/\lsun$.  This is consistent with clusters being
approximately isothermal, and being dominated by a relatively red
population of galaxies at all the redshifts observed.

The field luminosity density is calculated from the same data set so
that we obtain an $(M/L)_c$ in the same redshift range. We find
$(M/L)_c=1160\pm130h\msun/\lsun$, where the error is estimated from
field-to-field variations. This is consistent, within the errors, with
low redshift results, but it should not be compared directly to them
because the two are not derived in an identical manner. We conclude
that the cluster virial masses universally indicate that
$\Omega=0.25\pm0.05$ for clusters at $z\sim\onethird$, or
$\Omega_0=0.20\pm0.04$. Our measurement of $\Omega$ is done completely
within our data set which diminishes many of the possible selection
effects to small corrections to the luminosity, filter bands, and
shape of the luminosity function, since they are done identically for
the field and cluster galaxies. The concern that relative evolution of
field and cluster galaxies could lead to systematic errors is
partially constrained here and is addressed elsewhere (\cite{a2390}
and future papers). Cluster galaxies have largely ceased star
formation, so if anything, are likely to be less luminous that field
galaxies, meaning that the $\Omega$ given here is a mild overestimate.

Complementary to the mean $M_v/L$ value is the average $M_v/N_L$, which
is $3.5\pm0.4\times10^{12}h^{-1}\msun$ to $M_r(0)=-18.5$.  and
$5.7\pm0.7\times10^{12}$ to $M_r(0)=-19.5$ (excluding the two binary
clusters and MS1621+26).  The closure $M/N_L$ values at these two
luminosity levels give $\Omega=0.28\pm0.08$. Within the errors the
$M_v/L$ and the $M_v/N_L$ (for two different limiting magnitudes)
estimates for $\Omega$ are consistent with a single value, but,
because they are based on differing assumptions about the relative
evolution of the cluster and field galaxy population, the sum squared
errors in the two methods is used as an indicator of the systematic
error resulting from the luminosity normalization, finding that it is 0.09.

Our self-contained technique eliminates many of the systematic errors
of the $M/L$ and $M/N$ techniques for $\Omega$ estimation, but leaves
open the issue whether cluster galaxies are substantially more
clustered than the cluster mass. This will be addressed with the same
data set in a companion paper.

\acknowledgments
We thank all participants of the CNOC cluster survey for assistance in
obtaining and reducing these data. The Canadian Time Assignment
Committee for the CFHT generously allocated substantial grants of
observing time, and the CFHT organization provided the technical
support which made these observations feasible.  We gratefully
acknowledge financial support from NSERC and NRC of Canada.

\begin{table}
\caption{Cluster Redshift and Field Limits\label{tab:lim}}
\begin{tabular}{rrrrrr}
Name & $z$ & RA & Dec & $z_{lo}$ & $z_{hi}$ \\
 & & \multicolumn{2}{c}{\hmpc} & & \\
      A2390& 0.2280& 6.13& 1.11& 0.2175& 0.2395\\
MS0016$+$16& 0.5466& 2.19& 1.87& 0.5300& 0.5587\\
MS0302$+$16& 0.4246& 1.94& 1.73& 0.4160& 0.4300\\
MS0440$+$02& 0.1965& 3.32& 1.02& 0.1910& 0.2013\\
MS0451$+$02& 0.2011& 4.53& 1.57& 0.1919& 0.2120\\
MS0451$-$03& 0.5392& 2.26& 1.87& 0.5240& 0.5530\\
MS0839$+$29& 0.1928& 3.37& 0.96& 0.1820& 0.2024\\
MS0906$+$11& 0.1705& 1.06& 0.86& 0.1567& 0.1838\\
MS1006$+$12& 0.2605& 1.43& 1.20& 0.2530& 0.2680\\
MS1008$-$12& 0.3062& 1.61& 1.36& 0.2934& 0.3144\\
MS1224$+$20& 0.3255& 1.66& 1.35& 0.3180& 0.3310\\
MS1231$+$15& 0.2348& 1.33& 3.16& 0.2270& 0.2410\\
MS1358$+$62& 0.3290& 1.72& 4.27& 0.3150& 0.3420\\
MS1455$+$22& 0.2568& 1.47& 1.21& 0.2400& 0.2700\\
MS1512$+$36& 0.3726& 5.51& 1.70& 0.3640& 0.3800\\
MS1621$+$26& 0.4274& 2.04& 5.15& 0.4190& 0.4360\\

\end{tabular}
\end{table}

\begin{table}[h]
\caption{Velocity Dispersion Validation\label{tab:sig}}
\begin{tabular}{rrrr}
Name & $\sigma_1^\prime/\sigma_1$ & 1 s.d. interval\\
A2390 &  1.07 &  0.99 -  1.11\\
MS0016$+$16 &  1.07 &  0.93 -  1.15\\
MS0302$+$16 &  0.96 &  0.77 -  1.09\\
MS0440$+$02 &  0.95 &  0.60 -  1.14\\
MS0451$+$02 &  0.85 &  0.67 -  0.98\\
MS0451$-$03 &  0.96 &  0.84 -  1.01\\
MS0839$+$29 &  0.93 &  0.55 -  1.10\\
MS0906$+$11 &  0.37 &  0.22 -  0.49\\
MS1006$+$12 &  0.93 &  0.74 -  1.00\\
MS1008$-$12 &  1.03 &  0.90 -  1.09\\
MS1224$+$20 &  0.97 &  0.85 -  1.04\\
MS1231$+$15 &  1.04 &  0.95 -  1.10\\
MS1358$+$62 &  0.96 &  0.91 -  1.02\\
MS1455$+$22 &  0.97 &  0.85 -  1.09\\
MS1512$+$36 &  1.18 &  1.00 -  1.32\\
MS1621$+$26 &  0.92 &  0.83 -  1.00\\

\end{tabular}
\end{table}

\begin{table}[h]
\caption{CNOC Cluster Dynamical Parameters\label{tab:rsd}}
\begin{tabular}{rrrrrrrr}
Name & $z$ &N &$r_v$ & $\epsilon_r$ & $\sigma_1$ & $\epsilon_\sigma$ &
	 $\overline{\rho}(r_v)$ \\
 & & &\multicolumn{2}{c}{\hmpc} & \multicolumn{2}{c}{\kms} & $\rho_c(z)$ \\
      A2390& 0.2280& 174& 1.972& 0.16& 1104&  63&  119\\
MS0016$+$16& 0.5466&  47& 0.978& 0.12& 1234& 128&  360\\
MS0302$+$16& 0.4246&  28& 0.502& 0.12&  639&  92&  442\\
MS0440$+$02& 0.1965&  36& 1.224& 0.16&  606&  62&   99\\
MS0451$+$02& 0.2011& 108& 1.371& 0.14& 1031&  73&  226\\
MS0451$-$03& 0.5392&  51& 0.880& 0.10& 1371& 105&  555\\
MS0839$+$29& 0.1928&  42& 0.558& 0.14&  756& 111&  745\\
MS0906$+$11& 0.1705&  78& 0.584& 0.04& 1888& 117& 4426\\
MS1006$+$12& 0.2605&  26& 0.632& 0.06&  906& 101&  738\\
MS1008$-$12& 0.3062&  67& 0.575& 0.05& 1054& 107& 1113\\
MS1224$+$20& 0.3255&  24& 0.478& 0.15&  802&  90&  903\\
MS1231$+$15& 0.2348&  73& 0.944& 0.09&  640&  65&  173\\
MS1358$+$62& 0.3290& 165& 1.472& 0.13&  934&  54&  128\\
MS1455$+$22& 0.2568&  49& 0.621& 0.04& 1133& 151& 1203\\
MS1512$+$36& 0.3726&  38& 1.019& 0.41&  690&  96&  136\\
MS1621$+$26& 0.4274&  98& 1.428& 0.15&  793&  55&   84\\

\end{tabular}
\end{table}

\begin{table}[h]
\caption{CNOC Cluster $M_v$ and $L$ \label{tab:ml}}
\begin{tabular}{rrrrrrrr}
Name & $\overline{\rho}(r_v)$& $M_v$ &$L$&$M_v/L$&$\epsilon_{M/L}$ &
	$M_{200}$ & $L_{200}$\\
 & $\rho_c(z)$ &$h^{-1}\msun$&$h^{-2}\lsun$  &
	\multicolumn{2}{c}{$h\msun/\lsun$} &$h^{-1}\msun$ & $h^{-2}\lsun$ \\
      A2390&  119& 1.7$\times 10^{15}$& 6.7$\times 10^{12}$& 248&  43&
1.3$\times 10^{15}$& 5.2$\times 10^{12}$\\
MS0016$+$16&  360& 1.0$\times 10^{15}$& 3.6$\times 10^{12}$& 288&  94&
1.3$\times 10^{15}$& 4.8$\times 10^{12}$\\
MS0302$+$16&  442& 1.4$\times 10^{14}$& 6.3$\times 10^{11}$& 224& 108&
2.1$\times 10^{14}$& 9.4$\times 10^{11}$\\
MS0440$+$02&   99& 3.1$\times 10^{14}$& 9.9$\times 10^{11}$& 312&  87&
2.2$\times 10^{14}$& 7.0$\times 10^{11}$\\
MS0451$+$02&  226& 1.0$\times 10^{15}$& 2.8$\times 10^{12}$& 357&  71&
1.1$\times 10^{15}$& 3.0$\times 10^{12}$\\
MS0451$-$03&  555& 1.1$\times 10^{15}$& 2.9$\times 10^{12}$& 393& 102&
1.8$\times 10^{15}$& 4.8$\times 10^{12}$\\
MS0839$+$29&  745& 2.2$\times 10^{14}$& 7.7$\times 10^{11}$& 285&  83&
4.2$\times 10^{14}$& 1.5$\times 10^{12}$\\
MS0906$+$11& 4426& 1.4$\times 10^{15}$& 1.8$\times 10^{12}$& 800& 181&
6.6$\times 10^{15}$& 8.5$\times 10^{12}$\\
MS1006$+$12&  738& 3.6$\times 10^{14}$& 1.2$\times 10^{12}$& 291&  92&
6.9$\times 10^{14}$& 2.3$\times 10^{12}$\\
MS1008$-$12& 1113& 4.4$\times 10^{14}$& 2.0$\times 10^{12}$& 220&  60&
1.0$\times 10^{15}$& 4.7$\times 10^{12}$\\
MS1224$+$20&  903& 2.1$\times 10^{14}$& 1.0$\times 10^{12}$& 212& 114&
4.5$\times 10^{14}$& 2.1$\times 10^{12}$\\
MS1231$+$15&  173& 2.7$\times 10^{14}$& 1.5$\times 10^{12}$& 176&  43&
2.5$\times 10^{14}$& 1.4$\times 10^{12}$\\
MS1358$+$62&  128& 8.9$\times 10^{14}$& 4.5$\times 10^{12}$& 197&  26&
7.1$\times 10^{14}$& 3.6$\times 10^{12}$\\
MS1455$+$22& 1203& 5.5$\times 10^{14}$& 9.3$\times 10^{11}$& 589& 201&
1.3$\times 10^{15}$& 2.3$\times 10^{12}$\\
MS1512$+$36&  136& 3.3$\times 10^{14}$& 1.4$\times 10^{12}$& 234& 156&
2.7$\times 10^{14}$& 1.2$\times 10^{12}$\\
MS1621$+$26&   84& 6.2$\times 10^{14}$& 4.1$\times 10^{12}$& 151&  37&
4.0$\times 10^{14}$& 2.7$\times 10^{12}$\\

\end{tabular}
\end{table}

\begin{table}[h]
\caption{CNOC Cluster $M/N$ \label{tab:mn}}
\begin{tabular}{rrrrr}
Name & $M_v/N(-18.5)$ &$\epsilon_{M/N}$ & $M/N(-19.5)$ & $\epsilon_{M/N}$ \\
 & \multicolumn{4}{c}{$h^{-1}\msun$} \\
A2390& 2.3$\times 10^{12}$& 3.6$\times 10^{11}$&	5.7$\times 10^{12}$&
1.1$\times 10^{12}$\\
MS0016$+$16& 5.8$\times 10^{12}$& 2.0$\times 10^{12}$&	5.8$\times 10^{12}$&
2.0$\times 10^{12}$\\
MS0302$+$16& 2.2$\times 10^{12}$& 8.6$\times 10^{11}$&	3.4$\times 10^{12}$&
2.1$\times 10^{12}$\\
MS0440$+$02& 3.3$\times 10^{12}$& 9.2$\times 10^{11}$&	6.8$\times 10^{12}$&
2.3$\times 10^{12}$\\
MS0451$+$02& 3.8$\times 10^{12}$& 5.9$\times 10^{11}$&	7.4$\times 10^{12}$&
1.7$\times 10^{12}$\\
MS0451$-$03& 7.6$\times 10^{12}$& 2.0$\times 10^{12}$&	7.6$\times 10^{12}$&
2.1$\times 10^{12}$\\
MS0839$+$29& 2.4$\times 10^{12}$& 7.6$\times 10^{11}$&	7.7$\times 10^{12}$&
3.2$\times 10^{12}$\\
MS0906$+$11& 9.6$\times 10^{12}$& 1.9$\times 10^{12}$&	2.2$\times 10^{13}$&
6.1$\times 10^{12}$\\
MS1006$+$12& 5.6$\times 10^{12}$& 1.8$\times 10^{12}$&	5.6$\times 10^{12}$&
1.8$\times 10^{12}$\\
MS1008$-$12& 2.6$\times 10^{12}$& 6.9$\times 10^{11}$&	5.0$\times 10^{12}$&
1.7$\times 10^{12}$\\
MS1224$+$20& 3.0$\times 10^{12}$& 1.6$\times 10^{12}$&	4.6$\times 10^{12}$&
2.1$\times 10^{12}$\\
MS1231$+$15& 1.8$\times 10^{12}$& 4.5$\times 10^{11}$&	4.1$\times 10^{12}$&
1.3$\times 10^{12}$\\
MS1358$+$62& 2.1$\times 10^{12}$& 2.6$\times 10^{11}$&	4.2$\times 10^{12}$&
6.2$\times 10^{11}$\\
MS1455$+$22& 5.7$\times 10^{12}$& 1.7$\times 10^{12}$&	1.4$\times 10^{13}$&
5.8$\times 10^{12}$\\
MS1512$+$36& 2.2$\times 10^{12}$& 1.5$\times 10^{12}$&	6.7$\times 10^{12}$&
4.4$\times 10^{12}$\\
MS1621$+$26& 2.0$\times 10^{12}$& 4.1$\times 10^{11}$&	2.6$\times 10^{12}$&
4.7$\times 10^{11}$\\

\end{tabular}
\end{table}

\begin{table}[h]
\caption{Average $M_v/L$ and $M_v/N_L$ varying the Absolute Magnitude
Limit\label{tab:mlnav}}
\begin{tabular}{rrrrr}
$M_r(0)$ & $M_v/L$ & $\sigma_{M/L}$ \qquad & \qquad $M_v/N_L$
	& $\sigma_{M/N}$ \\
 & \multicolumn{2}{c}{$h\msun/\lsun$} & \multicolumn{2}{c}{$h^{-1}\msun$} \\
-18.0&  310&  113&$ 3.3\times10^{12}$&$ 1.7\times10^{12}$\\
-18.5&  283&   98&$ 3.5\times10^{12}$&$ 1.6\times10^{12}$\\
-19.0&  247&   80&$ 3.9\times10^{12}$&$ 1.7\times10^{12}$\\
-19.5&  241&   90&$ 5.7\times10^{12}$&$ 2.4\times10^{12}$\\
\end{tabular}
\end{table}

\begin{table}[h]
\caption{The Closure $M/L$ and $M/N_L$ Values varying the Absolute Magnitude
Limit
\label{tab:close}}
\begin{tabular}{rrrrr}
$M_r(0)$ & $\rho_c/j(>L)$ & $\sigma_{M/L}$ \qquad & \qquad $\rho_c/\phi(>L)$
	& $\sigma_{M/N}$ \\
 & \multicolumn{2}{c}{$h\msun/\lsun$} & \multicolumn{2}{c}{$h^{-1}\msun$} \\
-18.0& 1265&  149&$ 1.22\times10^{13}$&$ 2.1\times10^{12}$\\
-18.5& 1163&  130&$ 1.32\times10^{13}$&$ 1.9\times10^{12}$\\
-19.0& 1073&  104&$ 1.60\times10^{13}$&$ 1.5\times10^{12}$\\
-19.5& 1007&  103&$ 2.37\times10^{13}$&$ 2.6\times10^{12}$\\
\end{tabular}
\end{table}

\newpage
\figcaption[fig1.eps]{The redshift vs RA measured
from the brightest cluster galaxy direction for the full fields of the
16 clusters. The redshift increases from the vertex of the diagram,
and RA is converted to a transverse distance in physical
co-ordinates. \label{fig:raZ}}

\figcaption[fig2.eps]{Redshift vs Declination, as in Figure 1.\label{fig:dcZ}}

\figcaption[fig3.eps]{The redshift histograms in the vicinity
of the clusters. The dashed lines indicate the selected cluster
redshift range. \label{fig:zb}}

\figcaption[fig4.eps]{The sky locations of the galaxies
in the redshift range of the cluster. The circle area is proportional
to the galaxy's apparent magnitude. Each box is 1000 arcseconds on a
side.  The same layout of the clusters as in Figures 1, 2 and 3q is used.
North is to the top and East to the left.
\label{fig:cL}}

\figcaption[fig5.eps]{The corrected luminosities (top row) and
normalized colors (bottom row) for cluster galaxies (left column)
and field galaxies (right column). \label{fig:light}}

\figcaption[fig6.eps]{
The redshift corrected constant overdensity luminosity plotted against
the observed $\sigma_1$.
\label{fig:ls}}

\figcaption[fig7.eps]{The distribution of the deviations from
the mean $M_v/L$ (upper left, heavy lines for
the 6 well covered clusters), and the $M_v/L$ ratio as a function of
total blue fraction, color, mean interior density, velocity
dispersions and the redshift.
\label{fig:ml}}

\figcaption[fig8.eps]{The distribution of the deviations from
the mean $M_v/N_L$ (upper left, heavy lines for the 6 well covered
clusters) for $M_r(0)=-18.5$.  The $M_v/N_L$ ratio is plotted as a
function of the quantiies the previous figure.
\label{fig:mn}}

\end{document}